\documentclass[twocolumn,showpacs,preprintnumbers,amsmath,amssymb]{revtex4}
\usepackage{graphicx}% Include figure files
\usepackage{dcolumn}% Align table columns on decimal point
\usepackage{bm}% bold math

\begin{document}
%------------------------------------------------------------------------------
\title{Origin of the Universal Roughness Exponent of Brittle Fracture
Surfaces:\\ Correlated Percolation in the Damage Zone}

\author{Alex Hansen\footnote{Permanent Address: Department of Physics,
NTNU, N--7491 Trondheim, Norway. Email: Alex.Hansen@phys.ntnu.no.} 
and Jean Schmittbuhl\footnote{
Permanent Address: Departement de G{\'e}ologie, UMR CNRS 8538,
Ecole Normale Sup{\'e}rieure,
24, rue Lhomond, F--75231 Paris C{\'e}dex 05, France. Email: 
schmittb@geologie.ens.fr.}}
\affiliation{International Center for Condensed Matter Physics,\\
Universidade de Bras{\'\i}lia,\\
70919--970 Bras{\'\i}lia, Distrito Federal, Brazil}
\date{\today}
%--------------------------------------------------------------------
\begin{abstract} 
We suggest that the observed large-scale universal roughness of
brittle fracture surfaces is due to the fracture process being a
correlated percolation process in a self-generated quadratic damage
gradient.  We use the quasi-static two-dimensional fuse model as a
paradigm of a fracture model. We measure for this model, that exhibits
a correlated percolation process, the correlation length exponent $\nu
\approx 1.35$ and conjecture it to be equal to that of uncorrelated
percolation, 4/3.  We then show that the roughness exponent in the
fuse model is $\zeta=2\nu/(1+2\nu)= 8/11$.  This is in accordance with
the numerical value $\zeta=0.75$.  As for three-dimensional brittle
fractures, a mean-field theory gives $\nu=2$, leading to $\zeta=4/5$
in full accordance with the universally observed value $\zeta =0.80$.
\end{abstract} 
\pacs{83.80.Ab, 62.20.Mk, 81.40.Np}
\maketitle
%--------------------------------------------------------------------
Fracture surfaces in brittle materials show surprising scaling
properties \cite{b97}.  These were first seen in the mid-eighties
\cite{mpp84-bs85}.  They manifest themselves through self-affine
long-range height correlations.  That is, the conditional probability
density $p(x,y)$, i.e.\ the probability that the crack surface passes
within $dy$ of the height $y$ at position $x$ when it had height zero
at $x=0$, shows the invariance
\begin{equation}
\label{yhscaling}
\lambda^\zeta p(\lambda x,\lambda^\zeta y)=p(x,y)\;,
\end{equation}
where $\zeta$ is the roughness exponent.  In the early nineties
increasing experimental evidence hinted at the roughness exponent not
only existed, but had a {\it universal\/} value of about 0.80
\cite{blp90-mhhr92-sgr93-cw93-sss95}.  The experimental picture today
is even more complex: a) a second, smaller roughness exponent,
approximately equal to 0.5 has been observed on small length scales,
with a clear crossover length between the two regimes
\cite{dhbc96-dnbc97}; b) the growth of the roughness from an initial
straight notch shows anisotropy \cite{srb94} and a two-regime process
in case of quasi-brittle material such as wood \cite{mslv98}; c)
materials like sandstone for which the fracture is strongly
transgranular, show only a $\zeta=0.5$ self-affine scaling
\cite{bah98-m02}.  Simultaneously with these experiments, theoretical
and numerical work have been produced at a steady rate with the aim
of: (1) understanding why there is a self-affine scaling of the
roughness, (2) why there should be universality of the roughness
exponent and (3) how to unify observations and modelling
\cite{hhr91,bblp93-m93-bb94-ref97-bh98-rsad98-ran98-ljtz99-sra00-shh01-bhlp02}.

It is the aim of this letter to present a new possible explanation for
the observed universal roughness of brittle fracture surfaces at
larger scales.  We present our ideas using a paradigm of fracture
model: the quasi-static fuse model \cite{hr90}. 
Dynamical fuse models have been proposed and studied 
in the work of Sornette and
Vaneste \cite{sv92-lcs97}. The quasi-static fuse model consists of a
lattice where each bond is an ohmic resistor as long as the electrical
current it carries is below a threshold value.  If the threshold is
passed, the bond burns out irreversibly.  The threshold $t$ of each
bond is drawn from an uncorrelated distribution $p(t)$.  The lattice
is placed between electrical bus bars and an increasing current is
passed through it.  Numerically, the Kirchhoff equations are solved
with a voltage difference between the bus bars set to unity.  The
ratio between current $i_j$ and threshold $t_j$ for each bond $j$ is
calculated and the bond having the largest value, $\max_j (i_j/t_j)$
is identified and subsequently irreversibly removed.

In the limit of inifinite disorder --- i.\ e.\ when the threshold
distribution is on the verge of becoming non-normalizable, e.\ g.\
$p(t)\propto t^{-\alpha-1}$, where $1 \le t < \infty$ in the limit of
$\alpha\to 0$, the fuse problem becomes equivalent to a bond
percolation problem \cite{rhhg88}.  At more narrow disorders, a rich
phase diagram appears which is controlled by two parameters, the
exponent $\alpha$ which controls the threshold distribution tail 
towards infinitely large threshold values and the exponent $\beta$
which controls the tail of the threshold distribution towards zero:
$p(t)\propto t^{-1+\beta}$ where $0\le t \le 1$ \cite{hhr91a}.  For
smaller values of either $\alpha$ or $\beta$, the fuse model still shows
behavior very similar to percolation: The lattice stops conducting
after a finite percentage of bonds have burned out even when the
lattice size is extrapolated to infinity.  Close to breakdown,
critical exponents may be defined precisely as in the percolation
problem.  However, as the breakdown process in the fuse model is
highly correlated, there is no reason to expect these exponents to be
equal to those found in the percolation problem.  At even smaller
disorders, localization sets in.  

When the disorder is broad enough so that the fuse model behaves in a
percolation-like like manner, there is a diverging correlation length
$\xi \propto |p-p_c|^{-\nu}$, where $p$ is the density of broken bonds
and $p_c$ is the density at which an infinite lattice breaks down. For
classical percolation, $\nu=4/3$ \cite{sa92}.  For the fuse model away
from the infinite-disorder limit, $\nu$ has not been measured.  Three
scenarios are possible for the value of $\nu$: (1) $\nu$ depends on
the disorder.  Hence, it is not a universal quantity. (2) $\nu$ is
independent of the disorder but is different from 4/3.  In this case,
the fuse model defines a new universality class different from
standard percolation. (3) $\nu$ is the same in the fuse model as in
standard percolation.  Thus, the fuse model is in the universality
class of percolation.

\begin{figure}
\includegraphics[width=7cm,angle=270]{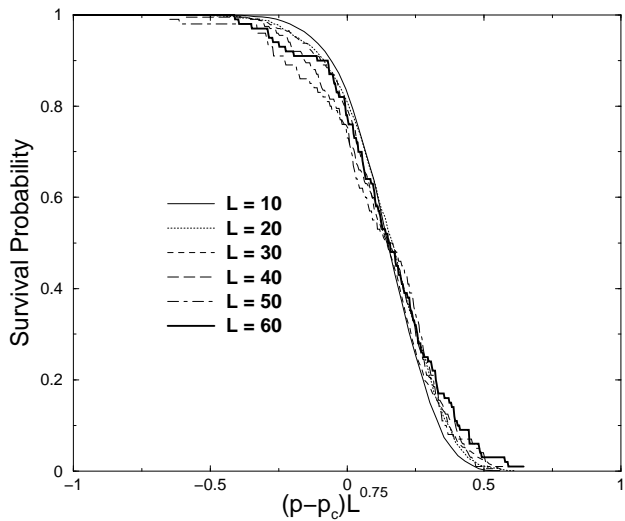}
\caption{Survival probability as a function of density of broken bonds
plotted against $(p-p_c)L^{1/\nu}$, where $p_c=0.3735$ and $\nu=4/3$ gives
a good data collapse for different lattice sizes.  The threshold
distribution was $p(t)\propto t^{-1+\beta}$ on the unit interval where
$\beta=1/10$.  The number of samples for each lattice size varied from 
2000 for $L=10$ to 80 for $L=60$.}
\label{fig1}
\end{figure}

\begin{figure}
\includegraphics[width=7cm,angle=270]{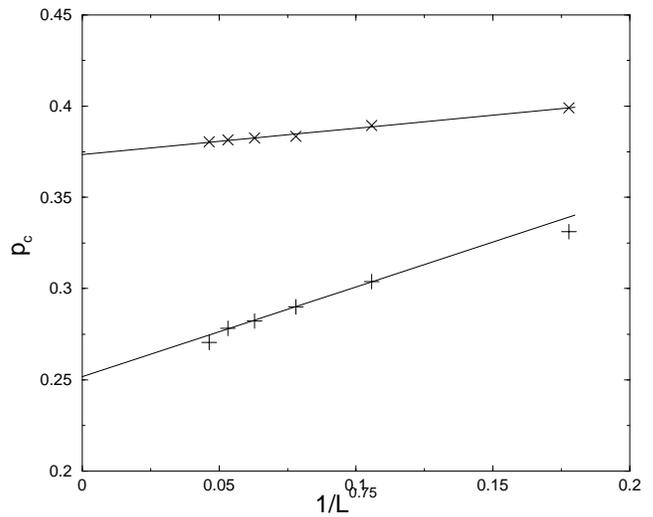}
\caption{50\% survival probability plotted against inverse lattice size
to the power 0.75 for threshold distributions on the unit interval, 
$p(t)\propto t^{-1+\beta}$ where $\beta=1/10$ ($\times$) and
$\beta=1/3$ (+).  The straight lines extrapolate to $p_c=0.3735$ and
$p_c=0.252$ respectively.}
\label{fig2}
\end{figure}

In order to determine which of these three scenarios is correct for
the two-dimensional fuse model, we studied the survival probability of
lattices for different system sizes and different disorders.  In
Fig.~\ref{fig1}, we show survival probability for the threshold
distribution $p(t)\propto t^{-1+\beta}$ when $0\le t \le 1$, where
$\beta=1/10$ as a function of density of broken bonds for different
lattice sizes. The collapse of the curves obtained for differents
sizes shows both that the survival probability is converging on a step
function at a finite $p=p_c$, and that an estimate of the coefficient
$1/\nu=0.75$. Indeed, we expect the survival probability to scale as
$L^{-1/\nu}$. In Fig.~\ref{fig2}, we confirm the estimate of $\nu$ by
showing the 50\% survival probability, as a function of lattice size
for this threshold distribution and for the threshold distribution
$p(t)\propto t^{-1+\beta}$ on the unit interval, where $\beta=1/3$.
Finite size scaling dictates that the effective density at which 50\%
of the lattices survive, $p_s$, behaves as
\begin{equation}
\label{50}
p_s=p_c-\frac{c}{L^{1/\nu}}\;,
\end{equation}  
where $c$ is a constant.  By adjusting
the value of $\nu$ until a straight line ensues, we determine the value
of $\nu$.  We find $\nu=4/3$ fits the data very well.  These results
are consistent with scenario (3) above: The two-dimensional 
fuse model is in the same universality class as classical two-dimensional
percolation.

If one tries to determine the scaling properties of the final crack
for disorders that are so broad that the system behaves as regular
percolation, the roughness exponent will be one, since the final crack
will be fractal with no anisotropy and so the width of the crack will
essentially be that of the lattice itself.  However, with more narrow 
threshold distributions, a non-linear gradient develops in the damage 
profile in the average current direction.  That is, if $y$ is the average 
current direction (which is normal to the two bus bars), then damage density
averaged in the orthogonal $x$ direction, $\langle p\rangle(y)$ takes
the form
\begin{equation}
\label{pgrad}
\langle p\rangle(y) = p_f-A \left(\frac{y-y_c}{l_y}\right)^2\;,
\end{equation}
where $A$ is a positive constant that depends on the width of the
threshold distribution and $l_y$ is the width of the damage
distribution.  The damage profile must surely be quadratic as the
system must be statistically mirror symmetric about $y_c$, where the
maximum damage occurs.  At breakdown, the maximum damage $p_f$ is
equal to the critical damage density $p_c$ and can be expressed in
terms of the correlation length~$\xi$: $(\langle p\rangle -p_c)\propto
\xi^{-1/\nu}$. As proposed by Sapoval {\it et al.\/} for percolation
in a gradient \cite{srg85}, we suggest to consider the region along
the damage zone that is at a distance corresponding to the correlation
length: $(y-y_c)\propto\xi$. Stating that the crack roughness is
proportional to the correlation length: $\xi\approx w$ and solving
Eq.\ (\ref{pgrad}) with the above conditions yield
\begin{equation}
w\sim {l_y}^{2\nu/(1+2\nu)}\;.
\end{equation}

The width of the damage profile, $l_y$, must be proportional to the
length of the system $L$.  The reason for this is that each broken
bond creates a disturbance in the average current field that enhances
the probability for a new bond to break in a finite-width cone 
which stretches out from each side of the bond in the direction 
approximately orthogonal to the average current direction.   Hence, as
long as the current enhancement is not sufficient to induce 
crack coalescence and create an unstable crack tip, the damage zone will spread
in the new cones in a random fashion.  This leads to 
\begin{equation}
\label{lyl}
l_y \propto L\;.
\end{equation}  

\begin{figure}
\includegraphics[width=7cm,angle=270]{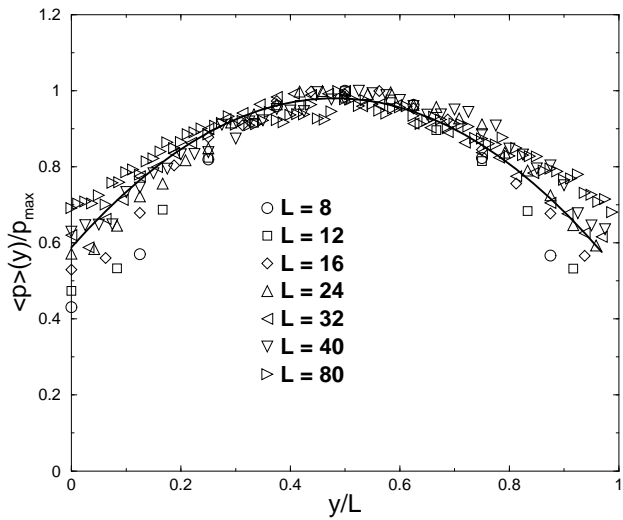}
\caption{Damage profile $\langle p\rangle (y)$ normalized so that its
its maximum is set to unity plotted against $y/L$ for different lattice
sizes $L$ and for the threshold distribution $p(t)\propto t^{-1+\beta}$
on the unit interval and where $\beta=1$. The curve is a quadratic best 
fit based on the $L=32$ data in accordance with Eq.\ (\ref{pgrad}).}
\label{fig3}
\end{figure}

\begin{figure}
\includegraphics[width=7cm,angle=270]{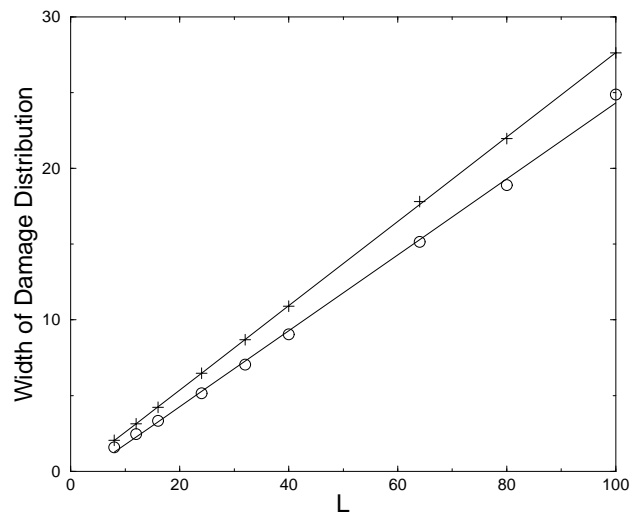}
\caption{Width of the damage distributions shown in Fig.\ \ref{fig3}
(+) and one based on the threshold distribution $p(t)\propto t^{-1+\beta}$
on the unit interval, whith $\beta=1/3$ ($\circ$) plotted against $L$.  
The straight lines are linear fits to the data.}
\label{fig4}
\end{figure}

In Fig.~\ref{fig3}, we show the damage profile averaged 
over many samples and for many lattice sizes plotted against $y/L$.  We note
from Fig.\ \ref{fig3} that the profiles clearly follow
Eq.\ (\ref{pgrad}). The collapse of the damage profiles shows that they
are functions of the combination $y/L$ and $L$ does not enter in any
other way. This results is confirmed in Fig.\ \ref{fig4} where the width
of the damage zone $l_y$ is plotted versus the system size $L$ for two
different threshold distributions. Both show a good linear
behavior in accordance with Eq.\ (\ref{lyl}).
Hence, the width of the crack scales as
\begin{equation}
\label{grad2}
w\sim L^{2\nu/(1+2\nu)}\;.
\end{equation} 
We therefore conclude that the fracture rougness exponent of the fuse model is
\begin{equation}
\label{zeta}
\zeta=\frac{2\nu}{1+2\nu}=\frac{8}{11}\approx 0.73\;,
\end{equation}
where we have on the right hand side assumed that  $\nu=4/3$, the 
standard percolation value.  In Ref.\ \cite{hhr91}, $\zeta$ was measured
to be about 0.75 in the two-dimensional fuse model.  Hence, there is very
good agreement with Eq.~(\ref{zeta}).

We now extend our argument to the general case of quasi-static 3D
brittle fractures in heterogeneous materials. No measurements of $\nu$
exist for the brittle fracture problem.  However, a recent mean-field
theory obtained: $\nu =2$ \cite{tp02}, making $\nu$ very
different from the value found in standard three-dimensional
percolation, $\nu = 0.88$.  Using Eq.\ (\ref{zeta}), we arrive at
\begin{equation}
\label{zeta3}
\zeta=\frac{4}{5}\;,
\end{equation}
which is indeed in excellent agreement with the experimentally observed
roughness exponent for large scales, $\zeta=0.80$.

Why should this theory be applicable only to the large-scale exponent 
observed in brittle fracture, and not the exponent seen at small scales?
It is the large-scale exponent that describes the correlated behavior of
the damage field which finally will lead to the large-scale properties of
fracture surface.  The smaller exponent, which is close to 0.5, describes
opening of small cracks, and may be caused by corrugation waves propagating
elastically along the crack front \cite{bbfrr02}. At larger scales, where
a roughness exponent equal to 0.8 is observed, these waves are too weak to
influence the system.  At this larger scale, we propose that it is a 
correlated gradient percolation process which is responsible for the value of 
the roughness exponent. 

This work was partially funded by the CNRS PICS contract $\#753$ 
and the Norwegian Research Council, NFR. We thank Fernando Oliveira
and the ICCMP for financial support during our stay in Bras{\'\i}lia.
Discussions with K.\ J.\ M{\aa}l{\o}y and G.\ G.\ Batrouni are 
gratefully acknowledged.
%--------------------------------------------------------------------
% BIBLIOGRAPHY
% --------------------------------------------------------------------

% --------------------------------------------------------------------
\end{document}